\documentclass[aps,pre,reprint,superscriptaddress,showpacs]{revtex4-1}
\usepackage[utf8]{inputenc}
\usepackage{graphicx}
\usepackage{epstopdf}
\usepackage{amsmath}
\usepackage{amssymb}
\usepackage{color} 
\usepackage{soul}
\usepackage{placeins}

\graphicspath{{figs_v12/}}

\begin{document}


\title{Time-scale effects on the gain-loss asymmetry in stock indices}

\author{Bulcs\'u S\'andor} 
  \affiliation{Goethe University Frankfurt, Institute for Theoretical Physics, D-60438, Frankfurt am Main, Germany}
  \affiliation{Babe\c{s}-Bolyai University, Department of Physics, RO-400084, Cluj-Napoca, Romania}

\author{Ingve Simonsen}
  \affiliation{Norwegian University of Science and Technology, Department of Physics, NO-7491 Trondheim, Norway}  

\author{B\'alint Zsolt Nagy}
  \affiliation{Babe\c{s}-Bolyai University, Department of Economics, RO-400084, Cluj-Napoca, Romania}  
  
\author{Zolt\'an N\'eda }
  \email[]{zneda@phys.ubbcluj.ro}
  \affiliation{Babe\c{s}-Bolyai University, Department of Physics, RO-400084, Cluj-Napoca, Romania}

\date{\today}

\begin{abstract}

The gain-loss asymmetry, observed in the inverse statistics of stock indices is present for logarithmic 
return levels that are over $2\%$, and it is the result
of the non-Pearson type auto-correlations in the index. These non-Pearson type correlations can be viewed also as functionally dependent daily volatilities, extending for a finite time interval.  A generalized time-window shuffling method is used to show the existence of such auto-correlations. Their characteristic time-scale proves to be smaller (less than $25$ trading days) than what  was previously believed.  
It is also found that this characteristic time-scale has decreased with the appearance of  program trading in the stock market transactions. Connections with the leverage effect are  also established.

\end{abstract}

\maketitle

\section{Introduction}

Physicists and economists have been analyzing complex financial time 
series with forward statistics (see below) for many years~\cite{Bouchaud2000,Mantegna2000,Chakraborti2011}.  Inverse statistics has been introduced recently as an alternative way of describing the phenomena of turbulence~\cite{Jensen1999} and adapted to finance by Simonsen et al.~\cite{Simonsen2002} for analyzing stock market time-series. This approach is motivated by the fact that dynamics of turbulent fluids is similar to the behavior of stock markets: after longer resting periods abrupt bursts can appear intermittently~\cite{Mantegna1996,Mantegna2000}.  Although the intermittency between 
resting periods and burst was modeled with success also by stochastic volatility models~\cite{Fouque}, and such models were also appropriate  to approach several other stylized facts like: fat-tails and the leverage effect \cite{Perello2004,Perello2007,Eisler}, the use of inverse statistics allows for a further understanding of various single stocks and market indices, as well as foreign exchange data and even artificial markets~\cite{Simonsen2002,Jensen2003,Jensen2004,Zhou2005,Simonsen2007,Zaluska2006, Karpio2007,Lee2008,Grudziecki2008,Siven2009a,Ebadi2010}. The method of inverse statistics revealed an intriguing gain-loss asymmetry \cite{Jensen2003a}, that generated a lot of discussion concerning its origin and the time-scale of the auto-correlations in the index, responsible for this effect~\cite{Donangelo2006,Simonsen2007, Siven2009b,Balogh2010,Bouchaud2001,Ahlgren2007}. Here we address the time-scale problem by introducing a shuffled window method on the stock index. Our results offer new data that could be useful for a better understanding of the gain-loss asymmetry and suggest also some connections with the leverage effect~\cite{Cont2001,Ahlgren2007}. 

\section{The method of inverse statistics} 

The stock index with the longest history is the Dow Jones Industrial Average 
(DJIA),  therefore most of the statistical studies deals with this time-series. For our study we use primarily the daily closure prices of the DJIA index   
from 1 Oct 1928 to 1 Feb 2011.  This corresponds to more than 80 years of data, and more than 20000 trading days.  We emphasize however that the results presented for DJIA are quite general since they are confirmed also for the case of the S\&P500 and NASDAQ100 indices (see the supplementary materials).

The performance of stocks and markets over a certain time history is traditionally measured by the distribution of the  $r_{\Delta t}(t)$ {\em logarithmic return}~\cite{Simonsen2002}, 
which gives us the generated return over a certain time period $\Delta t$. 
For individual stocks and market indices it is defined as the logarithmic price change over a fixed  time interval, $\Delta t$:
\begin{equation}
 r_{\Delta t}(t)=s(t+\Delta t)-s(t)=\ln{\frac{S(t+\Delta t)}{S(t)}},
\end{equation}  
where $s(t)=\ln(S(t))$ denotes the {\em logarithmic index} ($S(t)$ denotes the value of the index or the price of a stock). 
 
The standard deviation of $r_1(t)$ daily log-returns is called (daily) \textit{volatility}. 
 For the DJIA index, the historical daily volatility of the log-returns is about $\sigma=0.011$, i.e. $\sigma\approx 1\%$~\cite{Sullivan2003}. 

Empirical results show, that the {\em distribution of logarithmic returns} can be approximated by a Gaussian distribution (typically for larger $\Delta t$), 
although there are important differences, such as the presence of fat tails~\cite{Mantegna1995,Mantegna2000,Bouchaud2000,Simonsen2002} (most pronounced for shorter $\Delta t$). 
The fat tails correspond to a much larger probability for large price changes than what is to be expected from  Gaussian statistics, 
an assumption made in the mainstream theoretical finance~\cite{Mantegna2000,Bouchaud2000,Jensen2003a}.
Similar results have been found using forward statistics for the study of fully developed turbulence in fluids. 
As a consequence, in several works these two, seemingly completely different phenomena,  were discussed in parallel~\cite{Mantegna1996,Bouchaud2000,Mantegna2000,Jensen2003}.

For a deeper understanding of the fluctuation processes  Simonsen et al. have investigated  in~\cite{Simonsen2002} the inverse question: what is the typical waiting time 
to generate a fluctuation of a given size in the price?
To answer this question, we have to determine for an index or a stock the distribution of $\tau_{\rho}$ time intervals 
needed to obtain a predefined return level $\rho$.  
Practically, if given a fixed logarithmic return target  $\rho$ (proposed by the investor) for a stock or an index, 
as well as a fixed investment date (when the investor buys some assets), 
by the inverse statistics the time span is estimated for which the log-return of the stock or index reaches for the first time the desired level $\rho$. 
This is also called the \textit{first passage time} through the level $\rho$~\cite{Simonsen2002,Redner2001}. 
In a mathematical formulation this is equivalent to:
\begin{equation}
 \tau_{\rho}(t)=\inf\{\Delta t\geqslant 0\,\,|\,\,  r_{\Delta t}(t)  \geqslant  \rho  \}, \text{ if  } \rho>0,
\end{equation} 
or
\begin{equation}
 \tau_{\rho}(t)=\inf\{\Delta t\geqslant 0\,\,|\,\,  r_{\Delta t}(t)  \leqslant  \rho  \}, \text{ if  } \rho<0.
\end{equation} 
The waiting time $\tau_{\rho}(t)$ is the momentary investment horizon for the proposed $\rho$ log-return value~\cite{Simonsen2002}, indicating the time interval an investor has to wait
if the investment was made at time $t$,  and he/she wants to achieve a $\rho$  log-return value at time $t+\tau_{\rho}$. In the literature, its time-averaged value is the \textit{investment horizon}. 
The normalized histogram of the first passage times for many $t$ starting times, gives the $p(\tau_{\rho})$ probability distribution of the momentary investment horizons.
The method described above is called the method of inverse statistics.
The distribution of the momentary investment horizons for the DJIA index in case of $|\rho|=5\sigma$  (i.e. $\approx 5\%$ return) is depicted in the left panel of Fig.~\ref{fig:p-tau_T1}.
The maximum of the distribution function determines the most probable waiting time for that log-return ($\tau^{*}_{\rho}$), or in other words the  \textit{optimal investment horizon} for that stock or index.
The distribution of the first passage times gives also information about the stochasticity of the underlying asset price~\cite{Biferale1999,Jensen1999,Perello2011}.

A simple Brownian motion approximation for the log-prices~\cite{Simonsen2002} would yield for the first passage time distribution 
 \begin{equation}
 p( \tau_{\rho})=\frac{|\rho|}{\sqrt{4 \pi  D  \tau_{\rho}^3}} \exp{[-\rho^2/4D\tau_{\rho}]} 
 \label{fptd}
 \end{equation}
with $D$ a generalized diffusion constant. Since the first moment diverges, we determine the most probable first passage time: 
\begin{equation}
\tau^{*}_{\rho}=\frac{1}{6D}\rho^2\propto \rho^{\gamma},
\label{scaling-brown}
\end{equation} 
which should scale with an exponent  $\gamma=2$.
From this simple model, one also gets that the tail ($\tau_{\rho} \gg \tau^{*}_{\rho}$) of the distribution scales as
\begin{equation}
p(\tau_{\rho})\propto \tau_{\rho}^{-\alpha},
\end{equation}
with $\alpha=3/2$. Results for the DJIA~\cite{Simonsen2002} confirms this later scaling, however for $\gamma$ it yields a smaller exponent than the value expected for a Brownian motion. This is a clear sign that the daily volatilities are not independent variables, or in an other formulation one can state that some sort of 
time-like correlations are present in the dynamics of the index. Please note also from Eq.~(\ref{fptd}), that the distribution is symmetric relative to the sign of $\rho$, a result which is not confirmed by the data (see the next section).

\section{Gain-loss asymmetry}

In constructing the inverse statistics of the DJIA index also for the negative return levels (i.e. $\rho=-5\%$), 
it was found~\cite{Jensen2003a,Simonsen2007} that the distribution of investment horizons is similar in shape to the one for positive levels. 
However, there is one important difference: 
for negative return levels the maximum of the probability distribution is shifted to the left, 
generating about a $\Delta \tau_{\rho}\approx 13$ trading days difference in the optimal investment horizons. 
In the left panel of Fig.~\ref{fig:p-tau_T1} this asymmetry of the inverse statistics is presented for $\rho=+5\sigma$ and $\rho=-5\sigma$ log-returns.
It was found, that the asymmetry of inverse statistics is present for all the established stock indices, 
thus stock markets present a universal feature, called the \textit{gain-loss asymmetry}~\cite{Jensen2003a}. 
Contrary to indices, stock prices show a smaller degree of asymmetry~\cite{Donangelo2006,Siven2009b,Balogh2010}.
The asymmetry of the inverse statistics of stock markets is still a central problem of applied mathematics, 
econophysics and economics~\cite{Siven2009, Siven2009a, Ahlgren2007, Sandor2015}.

Minimal models have been proposed for explaining this intriguing fact. The fear-factor 
model~\cite{Donangelo2006,Simonsen2007} explains gain-loss asymmetry by a synchronization-like concept:  stronger stock-stock correlations during dropping markets than during market raises~\cite{Balogh2010}. 
Recently the idea of fear factor model was generalized by allowing longer time periods than one trading day for stock-stock correlations~\cite{Siven2009,Siven2009a}. 
By conducting a series of statistical investigations on the DJIA index and its constituting stocks, Balogh et.~al.~\cite{Balogh2010} have demonstrated, 
that indeed there is a stronger stock-stock correlation during periods of falling markets. This empirical result gives confidence both in the fear factor hypothesis~\cite{Donangelo2006,Simonsen2007}
and the generalized asymmetric synchronous market model~\cite{Siven2009a}.

Additional explanation for the gain-loss asymmetry is given by a simple one factor model~\cite{Bouchaud2001}, the Frustration Governed Market model~\cite{Ahlgren2007}, or the use of stochastic volatility models \cite{Siven2009a} (like EGARCH \cite{Nelson}). The problem has also been investigated in a thesis at the  
Swiss Federal Institute of Technology in Zürich~\cite{Lagger2012} under the guidance of Prof. Didier Sornette. 
Very recently~\cite{Sandor2015} an interesting analogy between the variations of stock indices and the dynamics of a one-dimensional spring-block model placed on a 
running conveyor belt was discussed. This simple mechanical system shows a similar gain-loss asymmetry in the inverse statistics, and also presents the leverage effect.
Although these works suggest the possibility that the gain-loss asymmetry and leverage effect might have the same origin based on the collective 
dynamics of the stocks, there is still a need for proving that the relevant time-scales of the underlying processes are the same for both effects.

\section{The shuffled time-window method}

Classical methods based on Pearson correlations are ineffective to prove the existence of auto-correlations in the index, a result which is in 
 agreement with the {\em efficient market hypothesis}~\cite{Fama1970}.  
  Lack of such first-order correlations do not exclude however the fact  that the daily returns and volatilities are not independent random variables. Such dependency can be viewed as higher order, or non-Pearson type correlations.

Here, by analyzing  the time-like variation of the stock 
index, we are using a statistical method that is suitable for giving further evidence for the presence of higher order/non Pearson type auto-correlations in the index and to determine their characteristic time-scale. The fact that daily volatilities are not independent variables are the reason for such non-Pearson type correlations. Stochastic volatility models \cite{asai} could account thus for their existence, however our aim here is not to model them. More specifically, our aim is to look for some special transformations applied to the time series of the index, that will result in the disappearance of the gain-loss asymmetry.

The problem is not as simple as it looks, since we are searching for a transformation,  which does not modify the distribution of daily returns of the considered index. 
Some methods that have been considered so far have altered the volatility of the daily returns. 
Using the wavelet transformation it has been suggested, that the asymmetry appears on time-scales longer than two months (between 64-128 trading days)~\cite{Siven2009}. This study concluded that by filtering out from the given time series fluctuations of periods larger than 64 days 
the inverse statistics of the index becomes symmetric again. It needs to be mentioned however, 
that in the method applied in~\cite{Siven2009} the values of daily returns are altered significantly, and as a result of this the distribution of returns and the volatility are also changed.  One can seriously question therefore the significance of this method.  
We have checked that using, instead of the wavelet transform, the
well-known Fourier transform, similar results are obtained.  Similarly with the wavelet transformation, the distribution of daily returns is 
again altered.

\begin{figure}[t!]
 \centering
 \includegraphics[width=0.95\columnwidth]{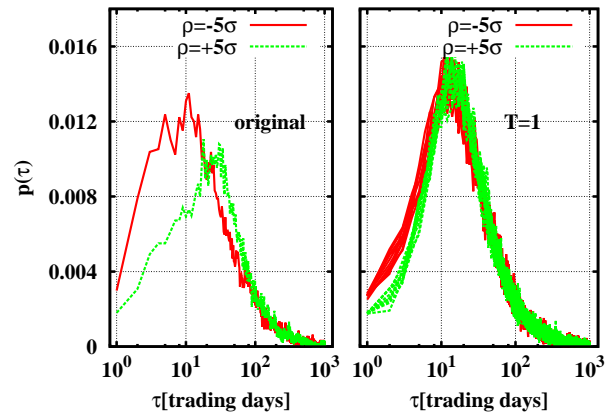}
 \caption{(Color online) Left: the investment horizon distributions of the original DJIA index. Right: the inverse statistics for the shuffled version of the index (is symmetric). The superimposed curves with the same color correspond to different permutations. The return level considered for the inverse statistics are five times larger than the daily volatility of returns: $\rho=5\sigma\approx 5\%$.}
 \label{fig:p-tau_T1}
\end{figure}

Recently a new idea was introduced to investigate the effect of correlations on this asymmetry~\cite{Siven2009a}.
Considering the time series of daily returns  and shuffling it randomly (permuting the elements one by one), 
the temporal dependence structure, thus correlations and connections of events (dependencies of the daily volatilities) can be destroyed, 
however the values of daily returns are kept unchanged. Therefore this method changes only the causality of events, 
but leaves all the other statistical information unchanged. An artificial index can be constructed from the time series generated in this manner, 
and the distributions of investment horizons can be investigated without altering the original information.
It was shown that the inverse statistics of this shuffled index becomes symmetric again (in the sense that the gain-loss asymmetry disappears)~\cite{Siven2009a}. Results in such sense are presented in Fig.~\ref{fig:p-tau_T1}. 
Since the shuffled index already produces a first-passage time distribution with a pronounced maximum (symmetric in $\rho$), it can be concluded that the existence of optimal investment horizons depends only on the distribution of daily returns, while the asymmetry is the combined consequence of the relative positions (the order in time) of these returns and their distribution. 
Generalized correlations routed in the fact that the daily volatilities are not independent increments are thus responsible for the gain-loss asymmetry. The source of this
 might be multiple, and we do not intend to study this problem here. 
It could be attributed for example to the cross-correlations, spanning several days,  between the stocks forming the index. As a result, the fear-factor model~\cite{Donangelo2006,Simonsen2007} still shows asymmetry after shuffling the returns, however this asymmetry disappears in the generalized asymmetric synchronous market model~\cite{Siven2009a}, where the falling stock prices stay synchronized for multiple days.

The window shuffling method can however be easily generalized, and  it seems suitable for detecting higher order auto-correlations in the index and to measure their relevant time-scales. 
In order to achieve this, we split the time series of daily returns into equally long time intervals consisting of $T$ trading days, 
called here \textit{time windows}. Then, we shuffle randomly these time windows without modifying the content inside any of them. 
By considering the permutations of these time windows instead of the permutations of the daily returns, 
we leave unmodified those events which have happened inside the time windows, but we break the correlation between those events that are 
separated by more than $T$ trading days. However, we should mention here that a part of the autocorrelations with a characteristic time scale
smaller than $T$ are also destroyed. More precisely $T_c/T$ part of the auto-correlations with characteristic time-length $T_c$ ($T_c<T$) are affected. Therefore
one would expect not a sharp but rather a smooth decay of the autocorrelations as one decreases $T$.  

\begin{figure}[t!]
 \centering
 \includegraphics[width=0.95\columnwidth]{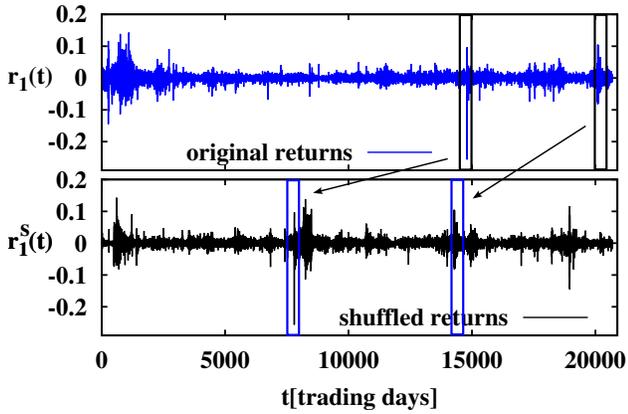}
 \caption{(Color online) Top panel: the original time series of daily returns. Bottom panel: the time series of the randomly permuted daily returns considering time windows of $T=500$ trading days. We illustrate the original and new positions of two selected time-windows.}
 \label{fig:r1-t_shuffled}
\end{figure}

As an example, if we split the original time series of the daily returns into intervals of $T=500$ trading days, we get $41$ time windows. In the top panel of Fig.~\ref{fig:r1-t_shuffled} we have marked two such time windows. Reshuffling randomly the order of all the windows, we get the time-series $r_1^s(t)$ from the figure in the bottom panel, where we have marked the 
new positions of the two selected time windows. 

\begin{figure}[t!]
 \centering
 \includegraphics[width=0.95\columnwidth]{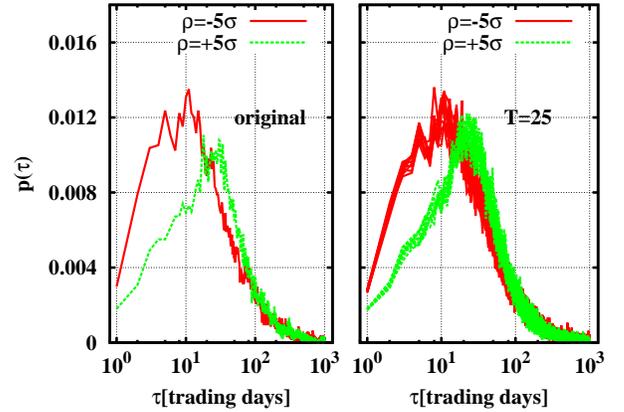}
 \caption{(Color online) Left: the investment horizons distributions of the original DJIA index. Right: the inverse statistics for the shuffled version of the index using time windows of $T=25$ trading days. The superimposed curves with the same colour correspond to different permutations. The return levels considered for the inverse statistics are five times larger than the volatility of returns: $|\rho|=5\sigma\approx 5\%$.}
 \label{fig:p-tau_T25}
\end{figure}

From Fig.~\ref{fig:p-tau_T1} we learned that for $T=1$, the gain-loss asymmetry disappears. Considering now 
time windows of $T=25$ trading days and the same $|\rho|=5 \sigma$ return level, we observe that the gain-loss asymmetry is almost as pronounced as in
the original time-series (Fig.~\ref{fig:p-tau_T25}). Therefore one can conclude that the auto-correlations causing the asymmetry for the $|\rho|=5\sigma$ return level manifest themselves on a time-scale shorter than $T=25$ trading days.
The same results can be found in the case of the SNP500 and the NASDAQ100 index (see the supplementary material).

\begin{figure}[t!]
 \centering
 \includegraphics[width=0.85\columnwidth]{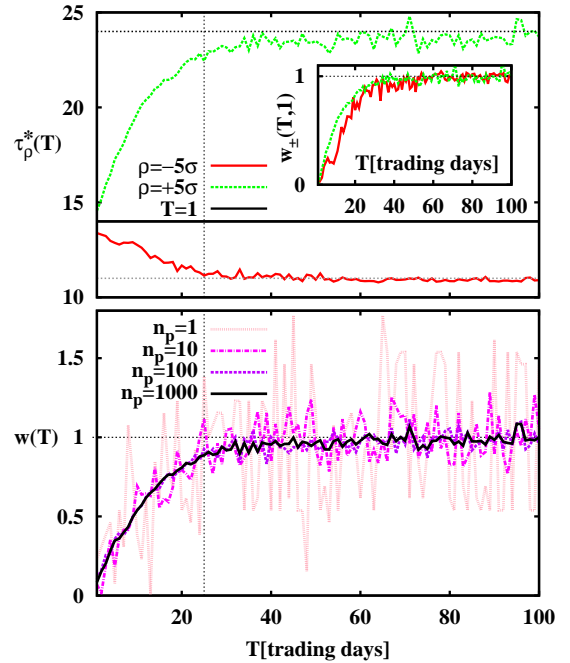}
 \caption{(Color online) Top panel: Positions of the $\tau^{\ast}_{|\rho|}(T)$ (green) and $\tau^{\ast}_{-|\rho|}(T)$ (red) maxima after shuffling with $T$ time window for $|\rho|=5\sigma$ logarithmic return level, in comparison to the fully shuffled case, $\tau^*_{\rho}(1)=14$ days (black). The dashed black and grey horizontal lines are corresponding to $\tau^{\ast}_{|\rho|}=24$ and $\tau^{\ast}_{-|\rho|}=11$ days respectively, and are shown to guide the eyes. The inset shows the $w_{\pm}(T,1)$ dissimilarity parameters. Bottom panel: The corresponding asymmetry level $w(T)$ for different $n_p$ permutations, using  $\Delta\tau_\rho^*(\infty)=13$ days. The dashed vertical line denotes the $T=25$ shuffle window size, see also Fig.~\ref{fig:p-tau_T25}.}  
 \label{fig:w-T_np}
\end{figure}

In order to get a better understanding of the relevant time-scales of the non-Pearson type auto-correlations, and the difference of the index dynamics relative to a simple Brownian dynamics, we compute several quantities for a wide range of return levels ($\rho$) 
and time-window lengths ($T$).

As Fig.~\ref{fig:p-tau_T1} suggests, the $r_1(t)$ time series of the daily returns is organized in such a way, that the $\tau^*_{\pm|\rho|}$ positions of the maxima for positive/negative return levels are shifted to the right/left with respect to the $\tau^*_{\rho}(1)$ maximum of the fully shuffled index ($T=1$). As a result of the window shuffling these differences are decreased until they disappear as the $T$ size of the time-window is gradually decreased (see top panel of Fig.~\ref{fig:w-T_np}).
The position of the maxima ($\tau^{\ast}_{|\rho|}(T),\tau^{\ast}_{-|\rho|}(T)$) are determined as the most probable first passage time, i.e. $\tau_{\rho}^{\ast}(T)$ for which:

\begin{equation}
 p(\tau_{\rho}^{\ast}(T))>p(\tau_{\rho}(T)),  \qquad \forall\, \tau_{\rho}(T) \neq \tau_{\rho}^{\ast}(T).
\end{equation} 

 Note that the $T\rightarrow \infty$ limit gives the optimal investment horizon, thus we can use the notation $\tau_{\rho}^{\ast}(\infty)=\tau_{\rho}^{\ast}$.
One can define a parameter, characterizing the degree of dissimilarity of the index shuffled with a $T$ time-window and a fully shuffled index
\begin{equation}
  w_{\pm}(T,1)= \frac{\Delta \tau^{\ast}_{\pm|\rho|}(T,1)} {\Delta \tau^{\ast}_{\pm|\rho|}(\infty,1)},
\end{equation} 
where we have introduced the notation
\begin{equation}
  \Delta \tau^*_{\pm|\rho|}(T,1)=\tau^{\ast}_{\pm|\rho|}(T)-\tau^{\ast}_{\rho}(1),
\end{equation}

From Fig.~\ref{fig:w-T_np} we learn that the shuffled window method affects the position of 
both maxima $\tau^*_{\pm|\rho|}(T)$ relative to $\tau_{\rho}^*(1)$. We also learn from the inset that 
$w_{+}(T,1)$ and $w_{-}(T,1)$ have a similar trend: 
\begin{equation}
  w_{-}(T,1)\approx w_{+}(T,1).\
  \label{eq:w_plus_minus}
\end{equation} 
Analogously, one can also define a parameter measuring the degree of the asymmetry in the original time-series of the stock index as the relative time-difference of optimal investment horizons for positive and negative returns:
\begin{equation}
  w(T)= \frac{\Delta \tau^{\ast}_{\rho}(T)} {\Delta \tau^{\ast}_{\rho}(\infty)}\,,
\end{equation} 
with 
\begin{equation}
  \Delta \tau^*_{\rho}(T)=\tau^{\ast}_{|\rho|}(T)-\tau^{\ast}_{-|\rho|}(T).
\end{equation}
From  Eq.~(\ref{eq:w_plus_minus}) it results that the three parameters defined above are equivalent:
\begin{equation}
  w(T)\approx w_{-}(T,1) \approx w_{+}(T,1)  \label{scaling}
\end{equation} 
In the followings we will use the $w(T)$ parameter as a measure of asymmetry for convenience.

In order to improve the statistics, the results are averaged for $n_p$ different permutations. Furthermore, for each different permutation the partitioning is also redone, by choosing the first day randomly  from the original time series in the interval  $\{1,T\}$. In this manner, we can avoid the generation of the same partitioning of the daily returns in time windows of length $T$. Consequently, the log-returns corresponding to large stock market crashes will appear at different positions in the time windows for different permutations. By increasing the number $n_p$ of permutations, the curves get smoother. Results showing this trend are presented for $|\rho|=5\sigma$ return level on the bottom  Fig.~\ref{fig:w-T_np}. We observe that for $n_p=1000$ the fluctuations are reasonably low.

In the followings this value will be used for all the statistics. In order to be consistent, we perform always the same number of permutations, and the $\tau^*_{\pm|\rho|}$ values for the original index are computed by taking a large window-size, $T_{\infty}=1000$:
$\Delta \tau_{\rho}^*(1000) \rightarrow \Delta \tau^{\ast}_{\rho}(\infty)$, since $\tau^{\ast}_{\pm|\rho|}(1000) \rightarrow \tau^{\ast}_{\pm|\rho|}$.

A possibility to improve the accuracy of the results for $w(T)$ would be by determining the value of $\tau^*_{\pm|\rho|}(T)$ maxima from a proper fit for
the $p(\tau_{\rho})$ distribution. In order to proceed in such manner one first has to find a proper fitting form and than perform the presumably nonlinear fit. Such a more sophisticated method is beyond the scopes of the present study. Here we aim to give only a rough estimate for the characteristic time-scale of the non-Pearson type correlations that are responsible for the
gain-loss asymmetry.

\section{Characteristic time-scales}

\begin{figure}[h!]
 \centering
 \includegraphics[width=0.85\columnwidth]{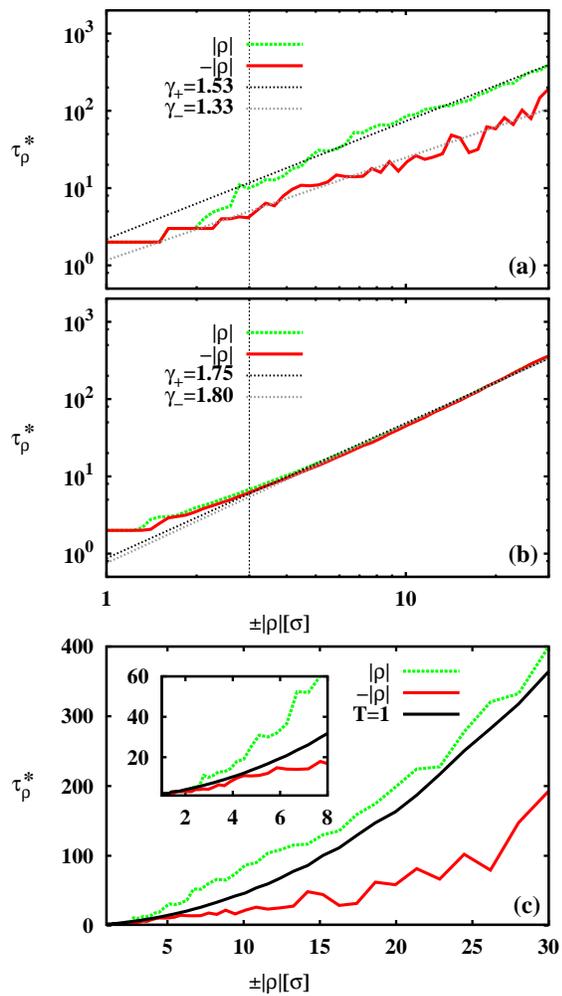}
 \caption{(Color online) The scaling of the $\tau_\rho^*$ optimal investment horizons for positive (green dashed curves) and negative (red continuos curves) $\rho$ return levels, using the original DJIA index (first panel) and the shuffled index (second panel) on log-log scales. The black dashed and gray continuous lines indicate the slopes fitted for $|\rho|>3\sigma$ return levels. The 
 $|\rho|>3\sigma$ limit is illustrated by the thin vertical line. Scaling exponents are given in the legend. We also present a comparative plot with an inset showing the region of small return levels using linear scales on the axis (third panel): $|\rho|\in[\sigma,8\sigma]$.}
 \label{fig:tau-rho_DJIA}
\end{figure}
   
First the averaged positions of the maxima in the inverse statistics, $\tau^{\ast}_{\rho}$, are studied for different 
return levels, $\rho$. We consider again both the case of the original index and the index shuffled with $T=1$ time-window. The results are plotted in Fig.~\ref{fig:tau-rho_DJIA}.

The results suggest some interesting conclusions. For the original index, we find that the gain-loss asymmetry is observable only for return rates over $2\sigma$, and it is increasing with increasing $\rho$ values.  For $T=1$ we kill all type of auto-correlations in the index, and as one would  naturally expect, the gain-loss asymmetry disappears for all return rates. The exponent $\gamma_{+}=1.8$ for the $\tau^{\ast}_{|\rho|}(\rho)$ scaling [Eq.~(\ref{scaling-brown})]  is much closer to the prediction of the simple Brownian dynamics ($\gamma=2$), in comparison with the original index where one gets: $\gamma_{+}=1.53$ for $\rho>0$ and $\gamma_{-}=1.33$ for $\rho<0$. This enables us to conclude that the index shuffled with $T=1$ gives a dynamics that is much closer to a simple Brownian motion, than the original one, but statistically still remains differences from the predictions of such a simple model. We believe the value of $\gamma<2$ for the shuffled index is a consequence of the empirical return distribution having fatter tails than the Gaussian distribution, a behavior that has also been demonstrated for synthetic time series (see the Appendix). From the third panel in Fig.~\ref{fig:tau-rho_DJIA} we can  conclude that the nature of the asymmetry depends strongly on the $\rho$ return level: for small return values $\rho\in[3\sigma,7\sigma]$ the dissimilarity is stronger for positive values, while for larger return levels the dissimilarity becomes more accentuated for negative returns.

\begin{figure}[t!]
 \centering
 \includegraphics[width=0.85\columnwidth,keepaspectratio=true]{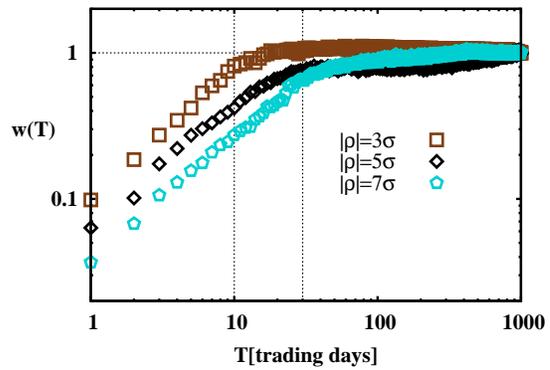}
 \caption{(Color online) Asymmetry remaining after shuffling with $T$ time window for different $\rho$ return levels in terms of the $\sigma$ volatility ($n_p=1000$).
}
 \label{fig:w-T_rho}
\end{figure}

Next, we investigate the dependence of the $w(T)$ asymmetry parameter as a function of the time-window length $T$, for those $\rho$ values where the gain-loss asymmetry is clearly observable in the original index:
($|\rho| \in[3\sigma,7\sigma]$). We did not considered return levels larger than $7 \sigma$, since in such case the statistics becomes poor. Results are plotted on Fig.~\ref{fig:w-T_rho}.  We learn from the figure that as the length of the  shuffling time window $T$ is decreasing  the amount of gain-loss asymmetry also decreases.  The results plotted on log-log scale show a  detectable cutoff value $T_c$, from where the auto-correlations in the index are strongly affected by the window shuffling method. This suggests the characteristic time-scale of the relevant auto-correlations, which turns out to be return level dependent. For higher return levels we find longer characteristic times in the $10-30$ days interval.  These limits are suggested by dashed lines in Fig.~\ref{fig:w-T_rho}.  

Similar result were found  for the NASDAQ100 and S\&P500 indices as well, even though the way they are computed from the component stock values is much different from the DJIA (see Fig.~\ref{fig:w-T_DJIA-SNP500-NASDAQ100} in Appendix). All these results suggest a much reduced characteristic time-scale for the non-Pearson type auto-correlations 
that are responsible for the gain-loss asymmetry than that reported in~\cite{Siven2009}. 

\begin{figure}[t!]
 \centering
 \includegraphics[width=0.85\columnwidth]{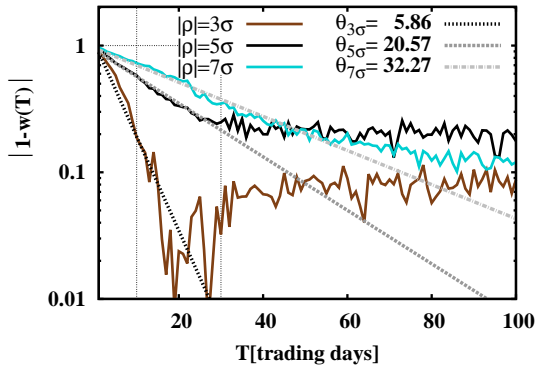}
 \caption{(Color online) The asymmetry measure of the DJIA index: $|1-w(T)|$ curves on a log-normal scale for different $\rho$ return levels in terms of the $\sigma$ volatility. Non-Pearson type correlations decreasing as $e^{-T/{\theta}}$ for $T<30$ days ($T_{\infty}=1000$).}
 \label{fig:w-T_exp}
\end{figure}

One could think of many other more sophisticated methods for determining the $T_c$ characteristic 
time-scale. One possibility would be to use the assumption that correlations are usually decaying in an exponential
manner, and consequently try an exponential fit of the type  $e^{-T/\theta}$ for $|1-w(T)|$. In this approximation $\theta$ would yield the $T_c$ characteristic time. As we illustrate in Fig.~\ref{fig:w-T_exp} such an exponential decay is indeed a reasonable
approximation in the limit of $T<30$ days. Moreover, the  fitted $\theta$ values (indicated in the figure) 
are in agreement with the visual estimates from Fig.~\ref{fig:w-T_rho}.

\begin{figure}[t!]
 \includegraphics[width=0.85\columnwidth,keepaspectratio=true]{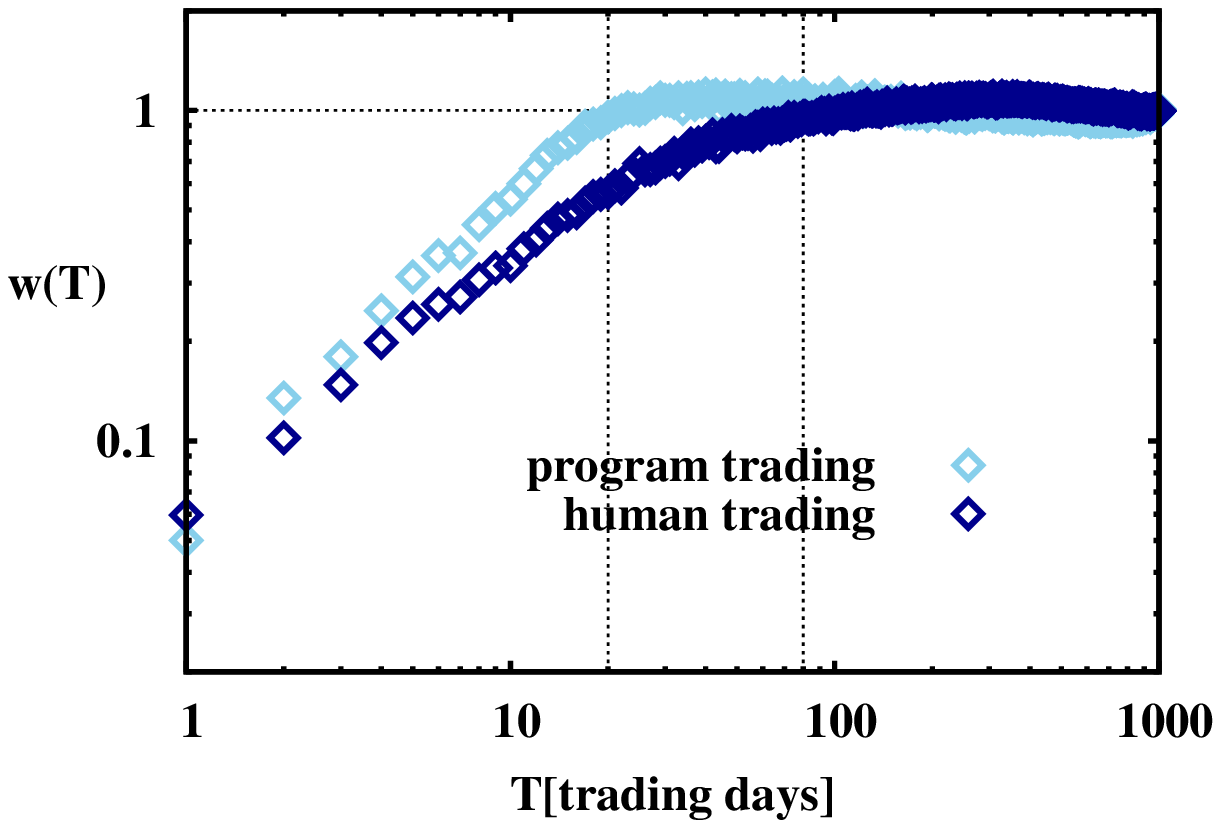}
 \includegraphics[width=0.85\columnwidth]{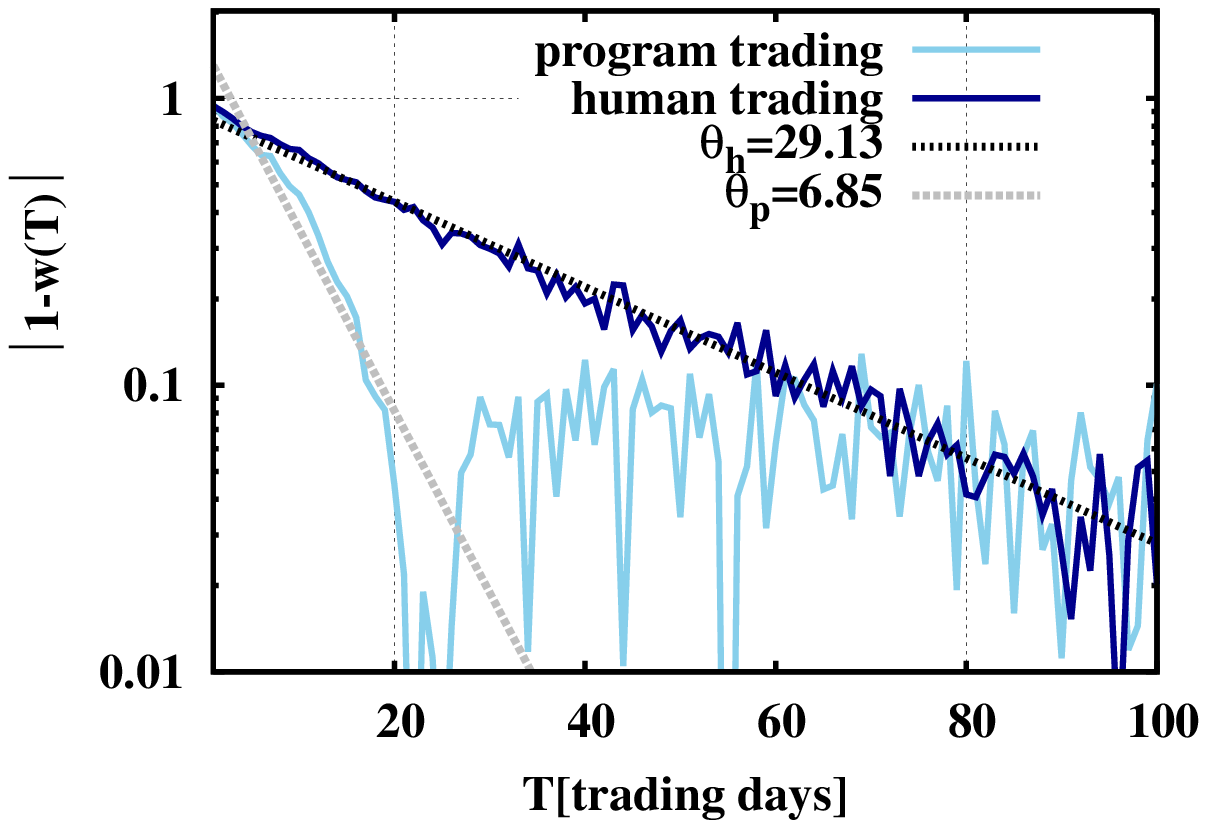}
 \caption{(Color online) Top: The asymmetry parameter of the DJIA index as a function of the $T$ time window 
 for human trading periods (1928-1980) and for program trading (1980-2011). 
 Bottom: The characteristic times $\theta_h$ and $\theta_p$ measured from the exponential fits corresponding to human and program trading, respectively. For both pictures the volatility of returns is chosen to be five times of the daily volatility: $|\rho|=5\sigma\approx 5 \%$ and $T_{\infty}=1000$.
 }
 \label{fig:w-T_hum-prog}
\end{figure}

Computer program controlled trading, called program trading, began in the 1970s and became widely used by the 1980s~\cite{Kolanovic2010}. The volume of asset transactions handled by computers did start to increase very rapidly in the early 2000s. 
Nowadays this volume  has surpassed  $40\%$ of the total trading volume. 
A reasonable hypothesis is that program trading changes the dynamics and the statistical characteristics of the stock market, since 
program trading, in contrast to human trading, is based on predefined algorithms.
To investigate the effect of program trading on the inverse statistics of the DJIA index, 
we have split the index into two parts: from 1928 to 1980 the period of mainly human trading, and from 1980 to 2011 the period where program trading is consistently present. 
Computing the asymmetry parameter as a function of $T$ for $|\rho|=5\sigma$, one gets the results presented 
on the top panel in Fig.~\ref{fig:w-T_hum-prog}.  One can clearly observe from the results of Fig.~\ref{fig:w-T_hum-prog} that the relevant time-scale responsible for the gain-loss asymmetry is significantly larger for the period 1928-1980 (about 30-80 days) 
than for the period 1980-2011 (about 10-20 days). For other return level values in the interval $|\rho| \in[3\sigma,7\sigma]$ the results are qualitatively similar.
The $\theta$ values determined from the exponential fit of the $|1-w(T)|$ function yields similar results (bottom plot of Fig.~\ref{fig:w-T_hum-prog}). For the program trading period
we get $\theta_p\approx 7$ days and for the human trading period $\theta_h\approx 30$ days.

 These results may  suggest that the introduction of program trading in the stock market transactions has a detectable influence on the stock indices, 
reducing the relevant time-scale of the non-Pearson type  auto-correlations that are responsible for the gain-loss asymmetry. This is a finding which has been raised by other authors as well (see Ref.~\cite{Ahlgren2007,Cont2001}). It is interesting to note here that    
the daily distributions of the returns remain similar for the human and program trading periods (see Fig.~\ref{fig:p-r2_prog-hum} in the Appendix).

\bigskip

\section{Discussion and Conclusions}

We have learned from the previous analysis that the gain-loss asymmetry is due to  non-Pearson type auto-correlations in the index. We also found that the characteristic time-scale of these auto-correlations are dependent on what we finally measure, but in any case, they are shorter than what was believed previously. Moreover, it is found that the characteristic time-scale decreased evidently by the
the appearance of program trading. The shuffled window method proved to be 
appropriate to detect these fine auto-correlations, which are not visible by using a simple Pearson correlation coefficient.

Here we speculate on connections with other stylized facts with similar time-scales.  
A well-known statistical property of financial time series is the \textit{leverage effect}, which states that 
the volatility of stocks (or index) tends to increase after price drops~\cite{Bouchaud2000,Cont2001,Ahlgren2007,Vargas2015}. 
One explanation of this effect is given by Bouchaud et al.~\cite{Bouchaud2000}: large daily drops, which increase volatility, are
often followed by rebound days. This could mean that some price drops are often exaggerated, 
due to the panic effect, and reach undervalued levels.  Gains are made in the following days when the asset price readjusts itself to its intrinsic value~\cite{Lagger2012}.  It should be mentioned that stochastic volatility models are also successful to explain the leverage effect (see for example \cite{Bouchaud2001,Perello2003pre,Siven2009a}). 

Mathematically the statement of the leverage effect can be quantified by the negative correlation between past returns and future volatility, and therefore it is measurable by the following correlation function~\cite{Ahlgren2007}:
\begin{equation}
  L_{\Delta t}(\tau)=\frac{\langle r^2_{\Delta t}(t+\tau)\cdot r_{\Delta t}(t)\rangle}{\langle r^2_{\Delta t}(t)\rangle^2}.
\end{equation}
For the DJIA index $ L_{\Delta t}(\tau) \approx 0$ for $\tau<0$, however for $0<\tau<25$ the normalized correlation function is negative with a minimum at $\tau=1$, 
and presents an exponential type relaxation~\cite{Lagger2012}. For $\tau>25$ the correlation relaxes to $0$. 
Note, that the falloff time of $ L_{\Delta t}(\tau)$ quantity is about $25$ days~\cite{Lagger2012}, which is on  similar time-scale with the auto-correlations responsible for the gain-loss asymmetry, presented above. 
Though the connection between the gain-loss asymmetry and the leverage effect  has been investigated already, 
to the best of our knowledge this similarity has never been pointed out up to now~\cite{Bouchaud2000,Ahlgren2007,Siven2009b,Lagger2012}.
 We believe that both effects are consequence of a fear-factor induced by the drop of the index, and seemingly relax on a similar time-scale. However, it has to be note that 
 presently it is believed that the gain-loss asymmetry may in principle exist with or without a leverage effect being present. For instance, within the model considered in Ref.~\cite{Ahlgren2007} this was demonstrated explicitly.
 
In an another line, we should note from Fig.~\ref{fig:w-T_np} that the $\Delta \tau^{\ast}_{\rho}(T)$ difference between the two maxima in the inverse statistics defines a kind of measure for the
length of the autocorrelations. As $T\rightarrow 1$ we get $\Delta \tau^{\ast}_{\rho}(T)\rightarrow 0$, suggesting a monotone relationship between the time-scale of the relevant auto-correlations 
and the strength of the gain-loss asymmetry. Moreover, as $T\rightarrow 1$ both of the maxima are shifting, $\tau_{|\rho|}^{\ast}$ shifting to the left and $\tau_{-|\rho|}^{\ast}$ shifting to right (see Figs.~\ref{fig:p-tau_T1} and \ref{fig:w-T_np}). Interestingly we get for $\Delta \tau^*_{\rho}(T)$ the same characteristic time scale
($\Delta \tau^*_{\rho} \in [10,30]$ days) as the one obtained for the relevant auto-correlation time by the window shuffling method. 
In this view, the relevant time-scale of auto-correlations we were searching for 
are already suggested by the positions of the maxima in Figs.~\ref{fig:p-tau_T1} and \ref{fig:w-T_np}.   

Finally, the main message from our study is that we revealed the characteristic time-scale on which the daily volatilities are functionally dependent (time-scale of the non-Pearson type autocorrelations in the index). Volatility models that aim to reproduce realistically the dynamics of the index, should take into account this characteristic time scale, and incorporate it in their assumptions.


\section*{Acknowledgement}
The work of B.S. was supported by the European Union 
and the State of Hungary, co-financed by the European
Social Fund in the framework of 
T\'AMOP 4.2.4.A/2-11-1-2012-0001 National Excellence Program.
B.S also acknowledges the support of the ``Collegium Talentum'' and the ``Sz\'ekely Forerunner Research Fellowship''. Z.N. acknowledges financial support from the Romanian "Idei" grant, No. PCE-IDEI-0348/2011. The research of I.S. was supported 
in part by The Research Council of Norway Contract No. 216699.

\newpage

\appendix
\section*{Appendix}

\begin{figure}
 \centering
 \includegraphics*[width=0.85\columnwidth]{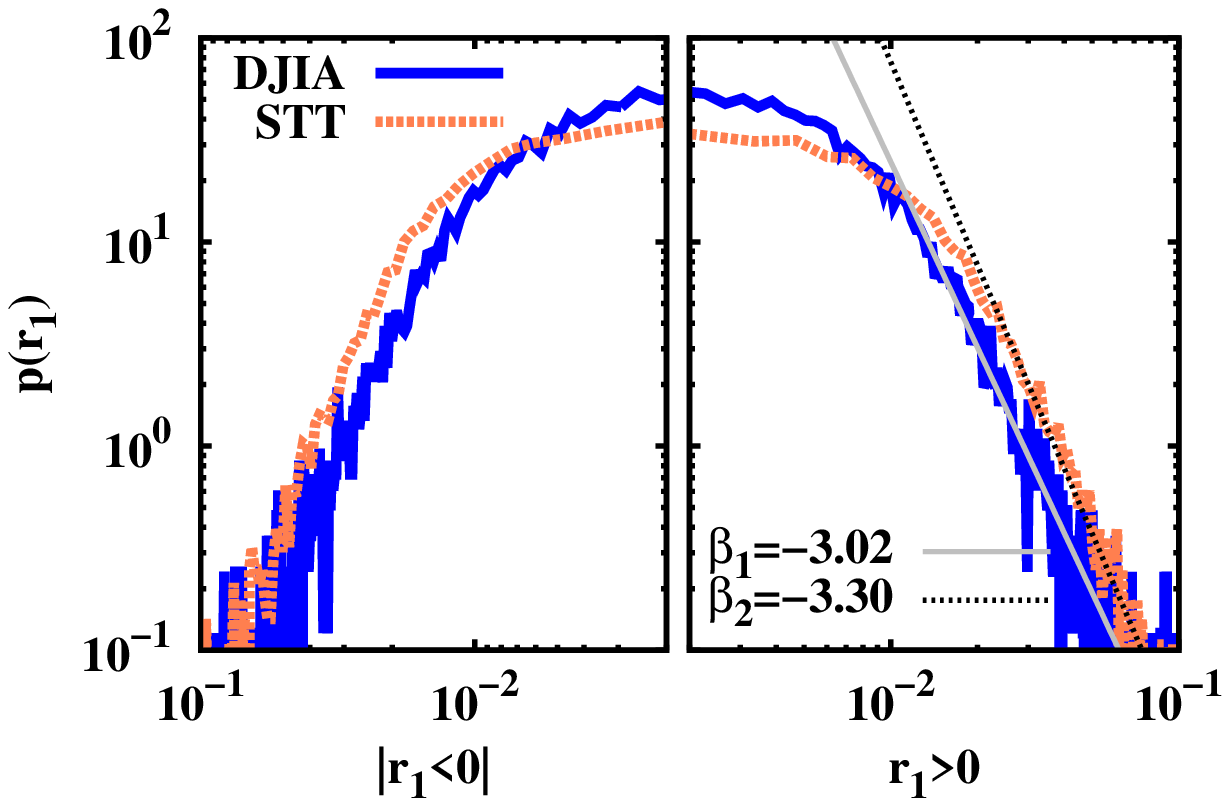}\\
 \includegraphics*[width=0.85\columnwidth]{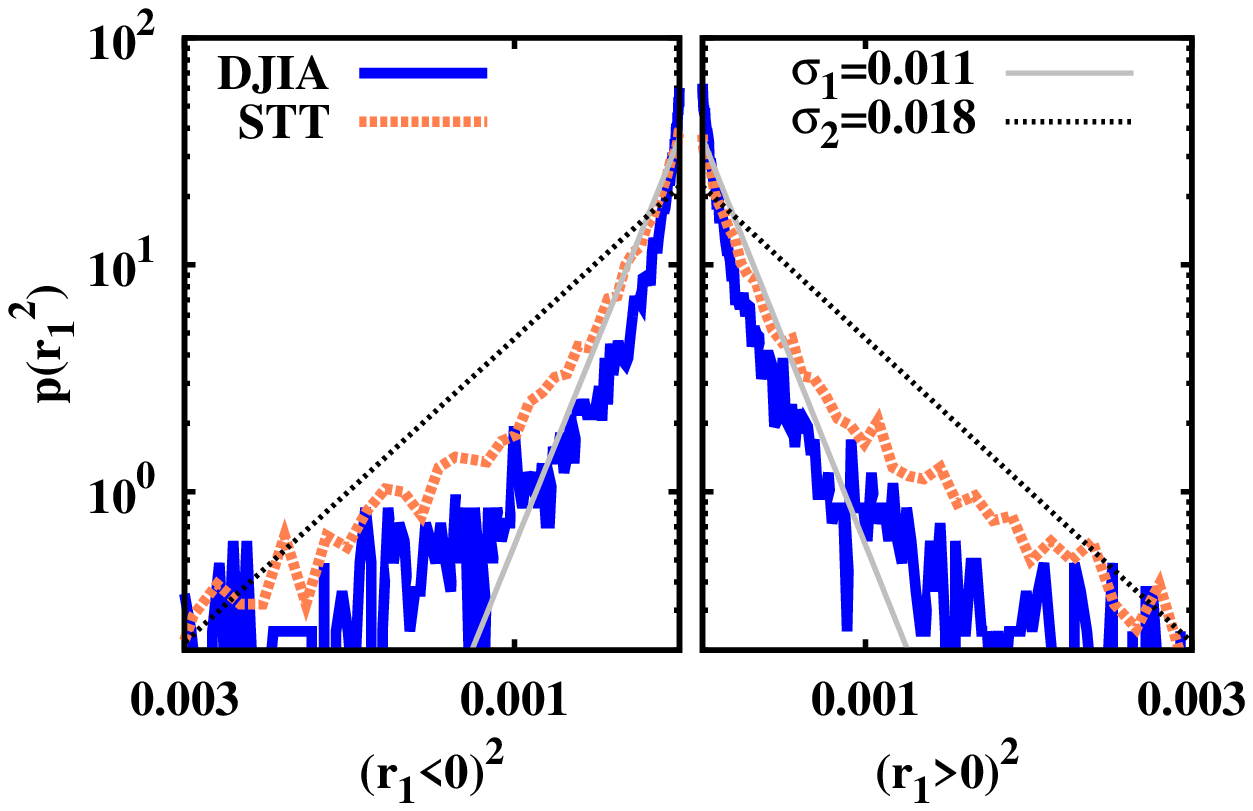}\
 \caption{(Color online) Normalized probability distributions of the positive and negative daily returns, $r_1(t)$, of the DJIA and the STT index. 
 \textit{Top}: Log-log scale, with the continuos gray and dashed black lines indicating two power-law fits 
 for the tail with exponents $\beta_1$ and $\beta_2$ respectively. Both distributions show a similar fat tailed. \textit{Bottom}:  Log-$x^2$ scale, in comparison with two Gaussian distribution functions $\mathcal{N} (0,\sigma_1^2)$ and $\mathcal{N} (0,\sigma_2^2)$ respectively.}
 \label{fig:p-r2_DJIA_STT}
\end{figure}

{\bf I.}
In order to generate a synthetic time series (daily returns) with fat tailed distribution and no autocorrelation, we have used Python's built-in Student's T distributed random number generator. 
Using $\nu=3$ shape parameter one can get non diverging second moments (note that the variance can not be smaller then $1$, getting actually $var = \nu/(\nu-2)=3$ in this case). By generating time series of the same length as our DJIA data, this would lead to the appearance of extremely large daily return values, and thus to a practically diverging index. To avoid this, we have rescaled these random variables, keeping the distribution normalized. This can be done easily by using the `scale` parameter of the Python package.  With a `scale=$0.01$` parameter the standard deviation of the generated times series (denoted by STT) is of the order of the volatility of the DJIA returns, $\sigma_{STT}=0.018$. 
The STT time series exhibits a non-Gaussian fat-tail distribution for the daily returns, similar with the one observed in the DJIA index~\cite{Chakraborti2011} (see  Fig.~\ref{fig:p-r2_DJIA_STT}).

\begin{figure}
 \centering
 \includegraphics[width=0.85\columnwidth]{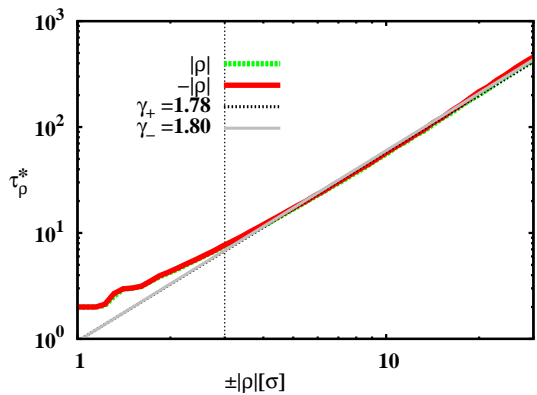}
 \caption{(Color online) The scaling of the $\tau_\rho^*$ optimal investment horizons for positive (green dashed curves) and negative (red continuos curves) $\rho$ return levels, using an artificial index with daily returns having a fat tailed Student's T distribution (STT). Please note the logarithmic scales. The black dashed and gray continuous line indicates the slopes  fitted for $|\rho|>3\sigma$ return levels, denoted by the thin vertical lines. The scaling exponents are given in the legend.}
 \label{fig:tau-rho_STT}
\end{figure}

Performing the inverse-statistics analysis on the STT index, we obtain similar $\gamma<2$ scaling exponents for the $\tau_\rho^*$ optimal investment horizons for positive  and negative  $\rho$ return levels as in the 
case of the shuffled DJIA index. From Fig.~\ref{fig:tau-rho_DJIA} we conclude $1.75<\gamma\leqslant1.8$ for the shuffled DJIA index while in the case of the STT we get  $1.78<\gamma\leqslant1.8$ (Fig.~\ref{fig:tau-rho_STT}).

\begin{figure}
 \includegraphics[width=0.85\columnwidth,keepaspectratio=true]{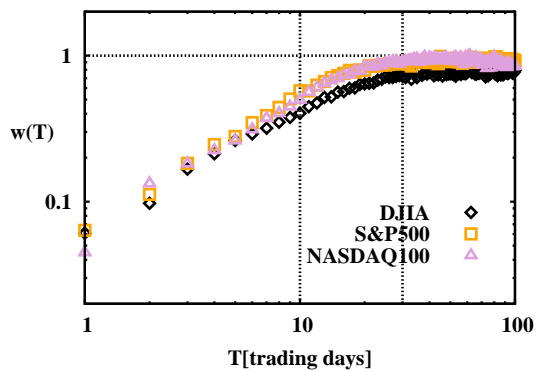}
 \caption{(Color online) The asymmetry parameter as a function of the $T$ time window for the DJIA, S\&P500 and NASDAQ100 indices. The volatility of returns are chosen to be five times of the daily volatilities of the respective indices: $|\rho|=5\sigma$. The curves start to saturate in the same interval: $T\in[10,30]$ trading days, indicated by dashed vertical lines.}
 \label{fig:w-T_DJIA-SNP500-NASDAQ100}
\end{figure}

{\bf II.} Considering the same type of analysis on the S\&P500 and NASDAQ100 indices as well, one can conclude that the underlying characteristic time-scales leading to the gain-loss asymmetry does not depend on the way the particular index has been constructed. To show this in Fig.~\ref{fig:w-T_DJIA-SNP500-NASDAQ100} we have plotted together for all the three indices the $w(T)$ asymmetry parameter as a function of the $T$ shuffle window size.  

\begin{figure}
 \centering
 \includegraphics*[width=0.85\columnwidth]{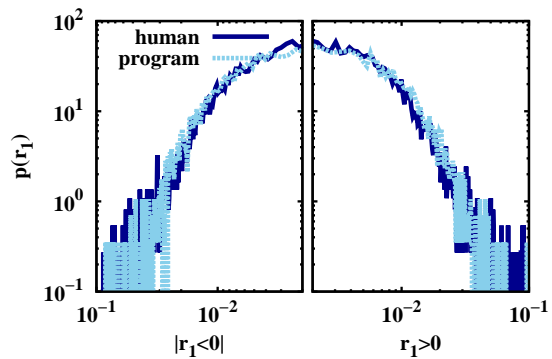}
 \caption{(Color online) Normalized probability distributions of daily returns, $r_1(t)$, of the DJIA index for the human and program trading periods.}
 \label{fig:p-r2_prog-hum}
\end{figure}

{\bf III.} Splitting the DJIA time-series in two periods, from 1928 to 1980 the period of mainly human trading, and from 1980 to 2011 the period where program trading is consistently present, we find that the distribution of the daily returns are not affected. Results in such sense are presented on Figure \ref{fig:p-r2_prog-hum}.

\FloatBarrier
\bibliographystyle{apsrev4-1}
\providecommand{\noopsort}[1]{}\providecommand{\singleletter}[1]{#1}%

\end{document}